\documentclass[twocolumn]{openjournal} 

\usepackage{orcidlink} 
\usepackage[T1]{fontenc}
\usepackage{ae,aecompl}

\usepackage{graphicx}	
\usepackage{amsmath}	
\usepackage{amssymb}	
\usepackage{natbib}

\usepackage{hyperref}
\hypersetup{
	colorlinks=true,
	urlcolor=blue,    
	linkcolor=black,  
	citecolor=blue,   
}

\defcitealias{Pittordis_2018}{PS18}
\defcitealias{Pittordis_2019}{PS19}
\defcitealias{Badry_2021}{ERH}
\defcitealias{Banik_2018_Centauri}{BZ18}

\hypersetup{citecolor=blue, 
            linkcolor=red, 
            menucolor=blue, 
            urlcolor=blue}  

\newcommand{\km}{{~\rm km}}
\newcommand{\s}{{~\rm s}}

\newcommand{\erg}{{~\rm erg}}
\newcommand{\yr}{{~\rm yr}}
\newcommand{\Myr}{{~\rm Myr}}

\newcommand{\pc}{{~\rm pc}}
\newcommand{\kpc}{{~\rm kpc}}

\begin{document}

\title{Comparing jet-shaped point symmetry in cluster cooling flows and supernovae}


\author{Noam Soker\,\orcidlink{0000-0003-0375-8987}} 
\affiliation{Department of Physics, Technion, Haifa, 3200003, Israel;  soker@technion.ac.il}

\begin{abstract}
I point out similarities between point-symmetric X-ray morphologies in cooling flow groups and clusters of galaxies, which are observed to be shaped by jets, and point-symmetric morphologies of eight core-collapse supernova (CCSN) remnants. I identify these similarities by qualitative eye inspection of multiwavelength images. I use these similarities to strengthen the jittering jet explosion mechanism (JJEM) of CCSNe, which predicts that the last pairs of jets to be launched by the newly born neutron star might shape some CCSN remnants to point-symmetric morphology. The point-symmetric morphologies in both types of objects are composed of two or more pairs of opposite bubbles (cavities), nozzles, some clumps, small protrusions (termed ears), and rims. The typically large volume of a CCSN remnant shaped by jets implies that the shaping jets carry an energy comparable to that of the ejecta, which in turn implies that jets exploded the remnant's massive star progenitor. The morphological similarities studied here add to the similarity of CCSN remnants, not only point-symmetric ones, to planetary nebulae shaped by jets. Together, these similarities solidify the JJEM as the main explosion mechanism of CCSNe. I consider the identification of point-symmetry in CCSNe, as expected by jet-shaping in the JJEM, to be the most severe challenge to the competing neutrino-driven explosion mechanism. I reiterate my earlier claim, but in a more vocal voice, that the main explosion mechanism of CCSNe is the JJEM. 
\end{abstract}


\section{Introduction} 
\label{sec:intro}

The thermal state of the hot gas in cooling flows in galaxies, in groups of galaxies, and clusters of galaxies is mainly determined by radiative cooling in the X-ray band and heating by the jets that the central active galactic nucleus (AGN) launches. In cooling flows, the radiative cooling time of the intracluster medium (ICM, or interstellar medium in galaxies) is shorter than the cluster's age. The mass cooling rate to low temperatures is much lower than the hot gas mass divided by the radiative cooling time (see however \citealt{Fabianetal2022}) because of the heating by jet-ICM interaction that operates in a negative cycle mechanism (for reviews see, e.g., \citealt{Fabian2012, McNamaraNulsen2012, Soker2016Rev, Werneretal2019}). The jet feedback mechanism operates (e.g., \citealt{BaumOdea1991}) as the hot ICM cools into cold clumps that feed the AGN  via an accretion disk, and this accretion disk launches jets that heat the ICM. 
  
The jets inflate X-ray deficient bubbles (cavities) in the ICM. Observations (Section \ref{sec:Comparing}) show that jets' launching occurs in episodes, each leaving a pair of opposite bubbles, and that the axis of the two opposite jets might vary from one jet-launching episode to the next.  Several theoretical mechanisms have been proposed to reorient the jets' axis, including the interaction of the accretion disk with the central supermassive black hole (e.g., \citealt{NatarajanPringle1998}), stochastic variations of the angular momentum of the accreted gas (e.g., \citealt{Babuletal2013}), an AGN driven by binary supermassive black holes  (e.g., \citealt{LiuFK2004, Gittietal2013}), and jittering where one jet-launching episode influence the next \citep{Soker2018Jit, Soker2022jit}. The typical reorientation timescales are few~$\Myr$  (e.g., \citealt{DennettThorpeetal2002, LalRao2007}).

The feedback cycle that operates by cold clumps that feed the AGN, rather than Bondi-type accretion of the hot gas, is termed the cold feedback mechanism or chaotic cold accretion (\citealt{PizzolatoSoker2005, PizzolatoSoker2010, Sharmaetal2012,  Gasparietal2013b, Gasparietal2015, Lietal2015, Prasadetal2015, VoitDonahue2015, Voitetal2015Natur, McKinleyetal2022, Ehlertetal2023}). 
Many recent papers support, in one way or another, the cold feedback mechanism 
(e.g., a partial list of papers from past five years, \citealt{Choudhuryetal2019, Ianietal2019, Roseetal2019, Russelletal2019, Sternetal2019, StorchiBergmann2019, Vantyghemetal2019, Voit2019, HardcastleCroston2020, Martzetal2020, Prasadetal2020, Eckertetal2021, Maccagnietal2021, Pasinietal2021, Qiuetal2021, Singhetal2021, Calzadillaetal2022, JimenezGallardoetal2022, Olivaresetal2022}). 
However, there are also claims for cases where Bondi accretion of hot gas feeds the central supermassive black hole, rather than of cold clumps (e.g.,  \citealt{Prasadetal2024}; see also discussion by \citealt{Bambicetal2023}). 
The reorientation of the jets' axis is likely related to the accretion process's stochastic nature onto the supermassive black hole in the cold feedback mechanism. Also, the multiphase ICM in the cold feedback mechanism influences the jets' propagation, e.g., as jets collide and are deflected by cold clouds. These properties are relevant to the goal of this study, which is to compare the morphologies of some cooling flows to some core-collapse supernova (CCSN) remnants.  

In a series of papers, I compared the morphologies that jet shape in the hot gas of cooling flows to jet-shaped morphologies in planetary nebulae (e.g.,  \cite{SokerBisker2006}\footnote{The images of cooling flows and planetary nebulae are in the arXiv version of the paper.}. While there are only a few tens of well-resolved morphologies in cooling flows, catalogs and surveys reveal several hundred well-resolved planetary nebula morphologies  (e.g., \citealt{Balick1987, Chuetal1987, Schwarzetal1992, CorradiSchwarz1995, Manchadoetal1996, Sahaietal2011, Parkeretal2016, Parker2022}). On the other hand, while the jets that shaped the observed planetary nebulae are not observed anymore in most planetary nebulae, the jets are well-resolved in cooling flows. Nonetheless, for more than three decades, studies have established the central role of jets in shaping planetary nebulae (e.g., \citealt{Morris1987, Soker1990AJ, SahaiTrauger1998, Sahai2000, Sahaietal2000, Sahaietal2007,  Boffinetal2012}), with many more studies in recent years (e.g., \citealt{Balicketal2020, GarciaSeguraetal2021, Guerreroetal2020, Guerreroetal2021,  Blackman2022, RechyGarciaetal2022, Bandyopadhyayetal2023, GomezMunozetal2023, Loraetal2023, Marietal2023, MoragaBaezetal2023, Uscangaetal2023, Kastneretal2024, Mirandaetal2024}).   

Several studies applied the knowledge from jet-shaping of planetary nebulae to argue that many CCSN remnants have been shaped by jets and that these jets played a role in the CCSN explosion mechanism (e.g., \citealt{Bearetal2017, GrichennerSoker2017,  BearSoker2018, Soker2022SNR0540, Soker2022RevRAA}). 
While the morphologies of planetary nebulae were compared separately to cooling flows and CCSN remnants, as far as I know, there has never been a systematic direct comparison of the morphologies of CCSN remnants with the morphologies of cooling flows in clusters and galaxies; this is the subject of this study. The main goal is to strengthen the claim that the CCSN remnants were shaped by jets and by that, to strengthen the jittering jets explosion mechanism (JJEM) according to which all CCSNe are exploded by jets. 
(e.g., \citealt{PapishSoker2011, PapishSoker2014Planar, GilkisSoker2014, GilkisSoker2015, GilkisSoker2016, Soker2020RAA, ShishkinSoker2021, ShishkinSoker2022, Soker2023gap, Soker2024Rev, WangShishkinSoker2024}).

In \cite{Soker2016Rev}, I review the operation of the jet feedback mechanism in several astrophysical classes of objects, including cooling flows, planetary nebulae, and CCSNe. However, I did not specifically compare the jet-shaped morphologies of the hot gas in cooling flows to CCSN remnants. 
The motivation to conduct this comparison now comes from the recent identification of point-symmetry in several CCSN remnants (for a review, see \citealt{Soker2024Rev}). In Section \ref{sec:Feedback}, I compare the basics of the jet feedback mechanism in CCSNe according to the JJEM with cooling flows. In Section \ref{sec:Comparing}, I find similarities between morphologies of CCSN remnants, particularly point-symmetrical morphologies, and morphologies of some cooling flows. Point-symmetrical morphologies are defined as having two or more opposite structural features to the center that do not share the same symmetry axis. I use these similar morphologies to strengthen the JJEM of CCSNe, as discussed in Section \ref{sec:Summary}.  In this study, the similarities are found by eye inspection. In future studies, we will start implementing more quantitative measures.  In Section \ref{sec:W49B}, I detour to examine the peculiar case of the W49B remnant, which is shaped by jets but might not be a CCSN.  

\section{Comparing the jet feedback mechanism} 
\label{sec:Feedback}

The jet feedback mechanism operates differently in CCSNe and cooling flows (for a review, see \citealt{Soker2016Rev}). In cooling flows, the jets heat the medium. Clumps cooling from the hot medium feed the central supermassive black hole (the cold feedback mechanism; Section \ref{sec:intro}). The jets are active for a long time, and the medium is optically thin, so the jets that heat the medium in cooling flows are observed, mainly in radio.  In the JJEM of CCSNe, the jets explode the star. They are active for a short time, a total jet-activity phase of $\simeq 1 - 10 \s$ or somewhat longer, much shorter than the dynamical time of the star. The jets interact inside the optically thick core and envelope of the star. For these, the jets are generally not observed (gamma-ray bursts are the exception).

 At the end of its evolution, the core of a massive star collapses to form a neutron star. According to the JJEM this neutron star launches the jittering jets as it accretes more mass from the collapsing core until the jets explode the star. In cases where more mass is accreted $\gtrsim 2.5 M_\odot$, the neutron star collapses into a black hole. In principle, the black hole can also launch jittering jets. However, according to the JJEM, black holes are formed when the jets remove the core mass inefficiently. This occurs when the specific angular momentum of the pre-collapse core is large such that the accretion disk around the neutron star maintains a more or less constant direction. The jets then do not expel gas from the equatorial plane and its vicinity, a mass that is accreted and turns the neutron star into a black hole \citep{Soker2023gap}. 

According to the JJEM (e.g., \citealt{PapishSoker2011, GilkisSoker2014, GilkisSoker2015, GilkisSoker2016}), there are $\approx {\rm few}$ to $\approx 30$ jet-launching episodes. The two opposite jets the newly born neutron star launches in each episode carry a mass of $\approx 10^{-3} M_\odot$ with an initial velocity of $\simeq 10^5 \km \s^{-1}$. The total explosion energy that jets power is typically $\simeq 10^{50} - 10^{52} \erg$. The direction of the jets varies more or less stochastically due to the stochastic convective motion in the pre-collapse stellar core (e.g., \citealt{ShishkinSoker2022, WangShishkinSoker2024}; or possibly in the envelope, e.g., \citealt{Quataertetal2019, AntoniQuataert2022, AntoniQuataert2023}). If pre-explosion core rotation exists, then the jets' axis variation is not completely stochastic (e.g., \citealt{Soker2023gap}). Also, the back reaction of the jets on the accreted mass acts to have the jets' axis in the same plane defined by the jets' axes of the two previous episodes \citep{PapishSoker2014Planar}.  Namely, after there are two not-aligned two jet-launching episodes, they define a plane. The axis of the next jet-launching episode tends to be in the same plane, and so are the axes of the following jet-launching episodes, as long as the angular momentum fluctuations of the accreted gas are not too large.   

Another critical difference is the activity time of jets.  Each jet-launching episode in the JJEM of CCSNe lasts for $\tau_{\rm jet} \simeq 0.01-0.1 \s$. This time is not much longer, or even shorter, than the relaxation time of the accreted material to a thin accretion disk. This implies that the two opposite sides of the accretion disk, which is thick rather than thin, are likely to differ in the exact structure and magnetic field properties \citep{Soker2024N63A}. The outcome might be,  according to the conjecture of the JJEM \citep{Soker2024N63A}, that the two sides will likely launch two opposite jets that differ in power. In turn, the jets will shape two opposite structural features different in their properties \citep{Soker2024N63A}. In cooling flows, most jets are active for much longer than the relaxation time of the accretion disk around the supermassive black hole, hence launching equal opposite jets.  

In cooling flows, the large-scale imprints of the jets are sub-sonic low-density hot bubbles, observed as X-ray deficient bubbles (cavities), in many cases filled with radio emission formed by the shocked jet's material. In CCSNe, the jets' imprints on the remnant are structural features in the, more or less, homologous super-sonic expanding ejecta. The jets are long gone and not active anymore. For that, in what follows, I apply observed jets in cooling flows to support the JJEM of CCSNe. Only the last one to several pairs of jets leave imprints in the CCSN remnant. Most early jets are choked inside the core and envelope of the exploding star and explode the star, leaving no large-scale morphological signatures (besides possibly in the nucleosynthesis process) that can be identified directly with jets.   
   
This study emphasizes point symmetry, which is a prediction of the JJEM. Only in the last two years have point-symmetric CCSN remnants been analyzed in the JJEM's frame. Identifying point symmetry in CCSN remnants is difficult because of several asymmetrical processes that occur during and after the explosion.
(1) The neutron star acquires a natal kick during the explosion process. 
(2) The explosion processes involve vigorous instabilities, some of which might deflect late jets. In the JJEM, the late jets shape the CCSN remnant because the early jets are choked inside the star and explode it. 
(3) Because the intermittent accretion disk that launches the jets has no time to relax, the two opposite jets it launches do not necessarily equal each other in power \citep{Soker2024N63A}. This implies that the two opposite structural features that the jets shape might differ in size, distance from the center of the explosion, and/or their exact shape.   (4) The hot ejecta at early times has a high sound speed, which acts to smooth high-pressure regions. (5) The ejecta will likely interact with a massive circumstellar material that the CCSN progenitor has lost over thousands of years pre-explosion (in winds and outbursts).  The circumstellar material can also possess point-symmetric structures. This is probably the cause of point-symmetry in type Ia supernova remnants (e.g., \citealt{Soker2024Rev}) that are descendants of exploding white dwarfs. However, circumstellar material cannot explain different compositions of structures, such as the jet in Cassiopeia A, cannot explain point-symmetry in the inner zones of the ejecta as seen in some CCSN remnants (e.g., the CCSN remnant 0540-69.3), and cannot account for signatures that suggest bending of jets, as in the Vela CCSN remnant.

Some similar processes also take place in cooling flows. 
($i$) The central AGN moves relative to the ICM. ($ii$) There are instabilities compatible with the cold feedback mechanism. Such instabilities that form dense clouds can deflect the jets. ($iii$) The ICM with which the jets interact might be asymmetrical due to the influence of cluster galaxies besides the cD galaxy that launches the jets. 

For all these processes, both in cooling flows and in CCSN remnants, the point symmetry might be far from pure and, in many cases, hard to identify. Nonetheless, in what follows, I find that the known activity of jets in cooling flows supports the claim for jet activity in CCSNe.  

\section{Morphologies in CCSNe and cooling flows} 
\label{sec:Comparing}


In this section, I compare the structures of seven CCSN remnants and one CCSN remnant candidate with identified point-symmetric morphology to some cooling flows. The observed jets in the cooling flows strengthen the claim that jets were active in shaping the ejecta during the explosion process of the respective CCSN remnants.  

\subsection{CCSN remnant N63A} 
\label{subsec:N63A}

\begin{figure}
\begin{center}
\includegraphics[trim=0.0cm 3.7cm 12.0cm 0.0cm ,clip, scale=0.76]{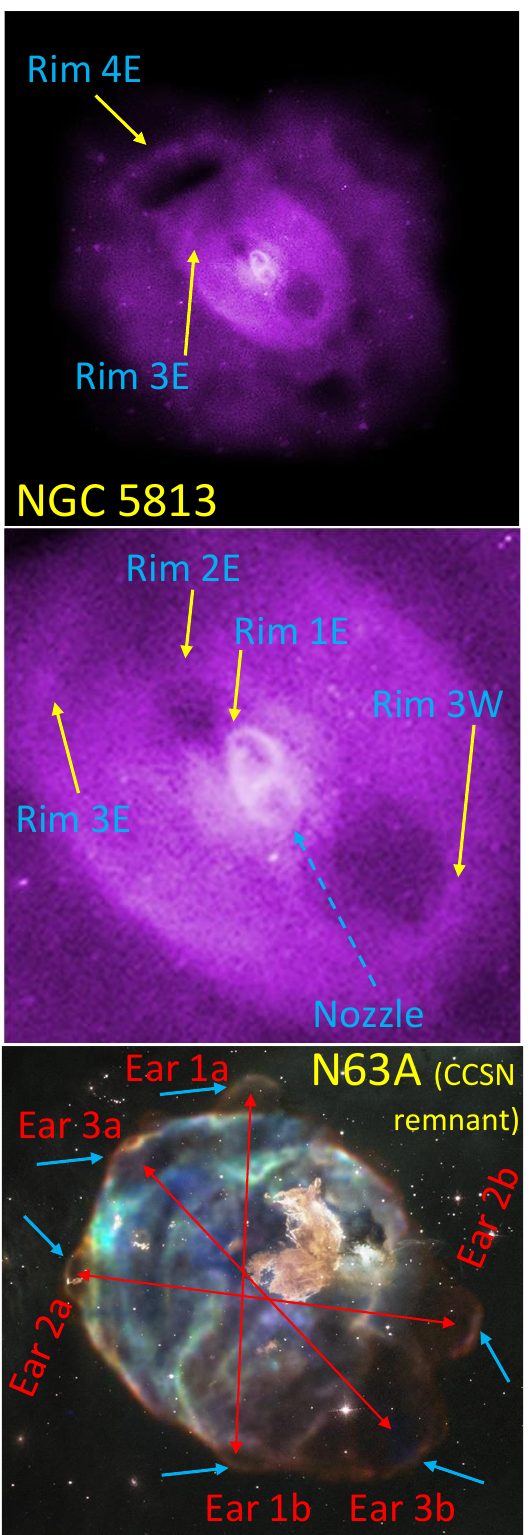} 
\end{center}
\caption{Upper panel: Chandra X-ray $80 \kpc \times 80 \kpc$ smoothed 0.3–3 keV image of the cooling flow galaxy group NGC 5813 adapted from Chandra site (credit: NASA/CXC/SAO/S, Randall) and based on \cite{Randalletal2015}. I added arrows pointing at the front rims of bubbles. 
Middle panel: The upper panel's inner $27 \kpc \times 27 \kpc$. 
Lower panel: A Chandra X-ray (red, green, blue for different X-ray energy bands) of CCSN remnant N63A; light brown zone is optical emission from Hubble. 
(Credit: Enhanced Image by Judy Schmidt based on images courtesy of NASA/CXC/SAO \& NASA/STScI.) I added light-blue arrows pointing at the rims of the six ears. Red double-headed arrows signify the point-symmetric structure. 
}
\label{Fig:NGC5813}
\end{figure}

In the two upper panels of Figure \ref{Fig:NGC5813}, I present X-ray images of the galaxy group NGC 5813.  
In the cooling flow NGC 5813 (e.g., \citealt{Randalletal2015}), the jets that inflated at least 7 X-ray deficient bubbles, or cavities, also carved a front rim, which appears as a front arc, in most of them, but not all. I mark the rims with solid yellow arrows on the upper two panels of Figure \ref{Fig:NGC5813}.    
However, there is no front rim in the closest southwest bubble. Instead, there is a nozzle, as I marked with the dashed-light-blue arrow. Also, there is no counter bubble to the second bubble in the northeast. Namely, there are no counter rims to Rim 1E and Rim 2E. I return to this asymmetry in section \ref{subsec:G107.7-5.1}   

Based on Chandra X-ray images of the CCSN remnant N63A, as the one in the lower panel of Figure \ref{Fig:NGC5813}, in particular on the new analysis by \cite{Karagozetal2023}, I argued recently \citep{Soker2024N63A} that the CCSN remnant N63A contains three pairs of ears that were shaped by three pairs of jets, the last jets among many more that exploded the progenitor of CCSN remnant N63A according to the JJEM. An ear is defined as a protrusion from the main part of the ejecta with a cross-section that monotonically decreases outward and has a different property from the main ejecta, like being fainter. With light-blue arrows, I point at the rim of each of the six ears of CCSN remnant N63A.  

While in CCSN remnants, jets that are related to the explosion process are not active anymore, in the case of NGC 5818, the radio emission that fills the bubbles \citep{Randalletal2011} indicates that jets inflated these bubbles and carved the rims. The similar rims in the galaxy group NGC 5813 and CCSN remnant N63A strengthen the claim that jets inflated the ears in N63A. 

\subsection{CCSN remnant SN 1987A} 
\label{subsec:SN1987A}

\begin{figure}
\begin{center}
\includegraphics[trim=0.0cm 11.1cm 6.0cm 0.0cm ,clip, scale=0.87]{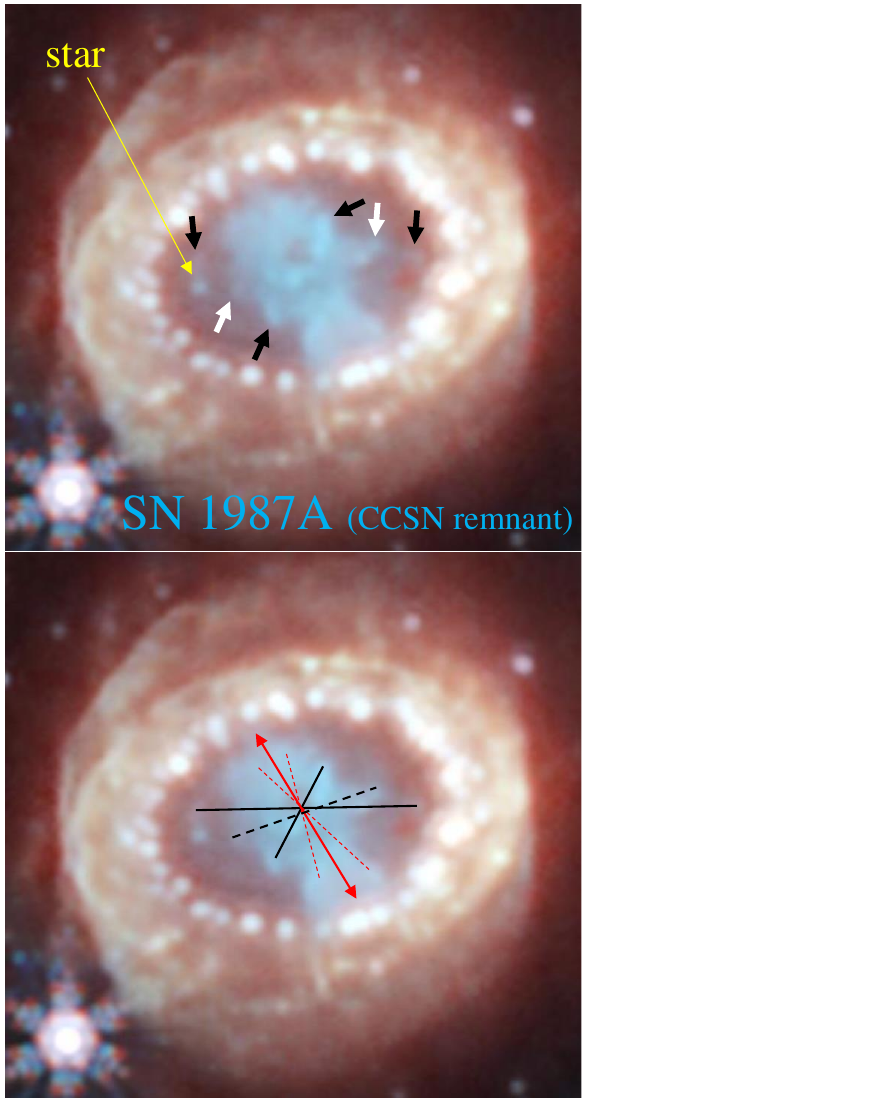} 
\end{center}
\caption{A JWST image of SN 1987A [source:\footnote{https://www.nasa.gov/missions/webb/webb-reveals-new-structures-within-iconic-supernova/}  NASA Webb Telescope Team, AUG 31, 2023; NASA, ESA, CSA: Matsuura, M. (Cardiff University), Arendt, R. (NASA’s Goddard Spaceflight Center \& University of Maryland, Baltimore County), Fransson, C.].
Colors represent wavelength (red: 4.44 microns; orange: 4.05 microns; yellow: 3.23 microns; cyan: 1.64 and 2.0 microns; blue: 1.5 microns). 
The black arrows in the upper panel and the western (on the right) white arrow are additions from \cite{Soker2024SN1987A}, and the white eastern arrow is a new addition. These arrows point at clumps, namely, some small areas in the outer regions of the blue-cyan structure that are brighter than their surroundings. The two solid-black lines (from \citealt{Soker2024SN1987A}) and the dashed-black line on the lower panel connect pairs of opposite clumps that likely were shaped by pairs of jets. The red double-headed arrow indicates the main axis of the CCSN remnant 1987A, with the two dashed-red lines that border a structure that might have been shaped by another pair of two opposite jets (from \citealt{Soker2024SN1987A}).  The thin-yellow arrow points at a bright knot that I considered a clump in \cite{Soker2024SN1987A} but seems to be a star. 
}
\label{Fig:SN1987A}
\end{figure}

Analyzing recent JWST observations of the remnant of SN 1987A that I present in Figure \ref{Fig:SN1987A}, I recently found hints of point-symmetrical morphological features in the very young remnant of SN 1987A \citep{Soker2024SN1987A}. I identified three pairs of clumps and a major symmetry axis. However, based on images from \cite{Arendtetal2023}, one clump that in \cite{Soker2024SN1987A} was identified as an ejecta clump seems to be a star; I point at it with a thin yellow arrow on the upper panel of Figure \ref{Fig:SN1987A}. Instead, the counter structure to the clump on the west might be another clump on the eastern side; I point at these two clumps with thick white arrows and connect them with a black-dashed line; this line misses by a small distance from the center of the other lines. I hold to my earlier claim that the clumps in the CCSN remnant of SN 1987A possess point symmetry, composed of the three pairs of clumps (three black lines on the lower panel on Figure \ref{Fig:SN1987A}) and the long axis (marked by the red lines).    

\begin{figure*}
\begin{center}
\includegraphics[trim=0.0cm 17.0cm 2.0cm 0.0cm ,clip, scale=0.83]{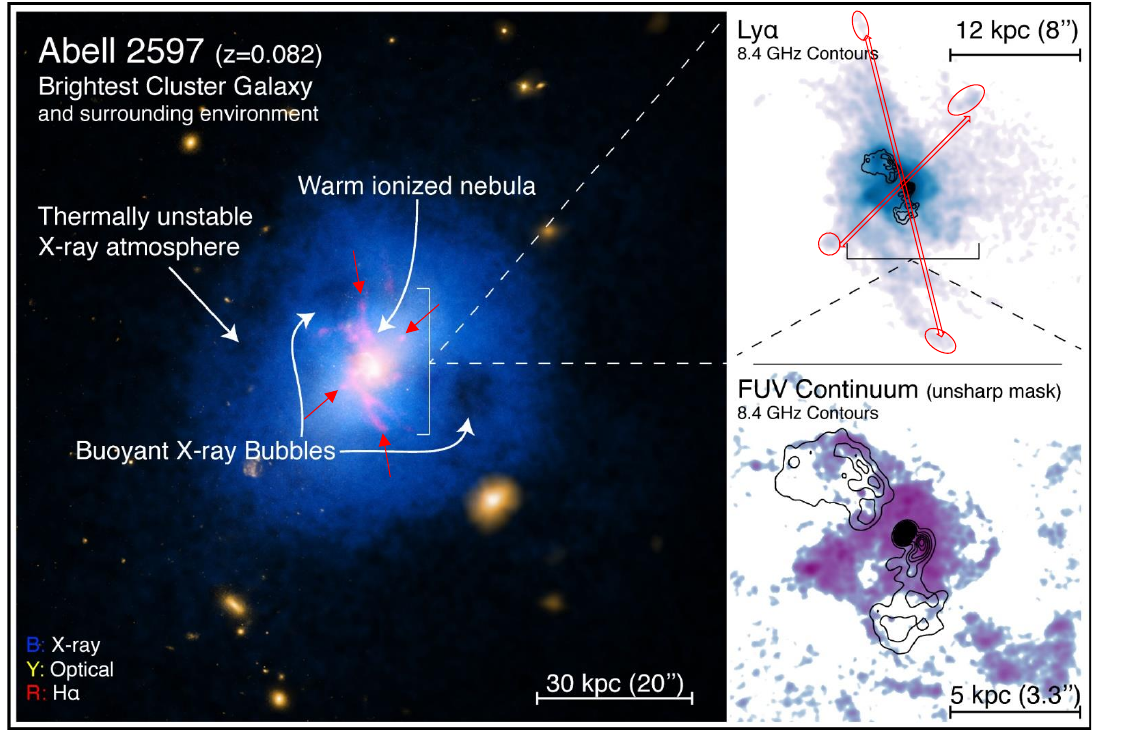}
\end{center}
\caption{Images of the cooling flow cluster Abell 2597 adapted from \cite{Tremblayetal2018}. Left panel: A composite image of Chandra X-ray in blue,  optical in yellow, and H$\alpha$+[N \textsc{II}] in red (Credit: X-ray: NASA/CXC/Michigan State University/G.Voit et al; optical: NASA/STScI and DSS; H$\alpha$: Carnegie Observatory/Magellan/W. Baade Telescope/University of Maryland/M.McDonald). 
Top right: HST/STIS MAMA image of Ly$\alpha$ emission associated with the ionized warm gas in the center. Overlaid are VLA 4.8 GHz radio contours. Bottom right: Unsharp mask of the HST/ACS SBC far-ultraviolet continuum image of the central regions of the remnant, with overlaid 8.4 GHz radio contours. East is to the left, and north is up. 
I added the red marks, four arrows in the left panel, and the double-headed arrows and ellipses in the top-right panel. These refer to two pairs of clumps. I use this point-symmetric clump morphology shaped by jets to argue for jet-shaped clumps in CCSN remnant 1987A (Figure \ref{Fig:SN1987A}).  
}
\label{Fig:A2597} 
\end{figure*}

The clumpy point symmetry is observed also in some cooling flows. In Figure \ref{Fig:A2597} I present images of Abell 2597 adapted from \cite{Tremblayetal2018}, who conducted a thorough study of the structure and kinematics of the multiphase ICM of the cooling flow cluster Abell 2597 (also \citealt{Tremblayetal2016}). I point at four clumps of warm ionized gas that seem composed of two pairs, as the double-headed-red arrows on the upper right panel indicate. The centers of the arrows are at their cross point. The structure of these clumps has been shaped by the jets and the bubbles they inflated. I consider this to support the suggestion that SN 1987A was exploded by jets and that possibly more than one pair of jets shaped the observed ejecta of CCSN remnant 1987A \citep{Soker2024SN1987A}.  

I comment here on the competing delayed neutrino explosion mechanism (e.g., \citealt{Fryeretal2022, Nakamuraetal2022, Olejaketal2022, Bocciolietal2023, Burrowsetal2023, BoccioliRoberti2024}, for some recent papers). According to the delayed neutrino explosion mechanism, clumps are formed by instabilities during the explosion process, as simulations show (e.g., \citealt{Wongwathanaratetal2015, Wongwathanaratetal2017, BurrowsVartanyan2021, Vartanyanetal2022, Orlando2023}). Similar instabilities are also expected in the JJEM. \cite{Orlandoetal2020}, for example, simulated the morphology of SN 1987A in the frame of the delayed neutrino explosion mechanism. However, the delayed neutrino explosion mechanism does not account for point-symmetric morphologies.  The reason is that there is no mechanism to define symmetry planes with varying directions; each plane includes two opposite outflows having similar properties. The only meaningful symmetry plane with a symmetry axis is the equatorial plane of the pre-collapse core rotation. Namely, the delayed neutrino mechanism might account, in principle, for one pair of structures, like one pair of ears,  although it is not expected. However, the delayed-neutrino explosion mechanism does not explain two or more opposite pairs. The JJEM predicts such point-symmetric morphologies.  

\subsection{CCSN remnant G321.3–3.9} 
\label{subsec:G321.3–3.9}

\begin{figure}
\begin{center}
\includegraphics[trim=1.0cm 6.0cm 4.0cm 0.2cm ,clip, scale=0.53]{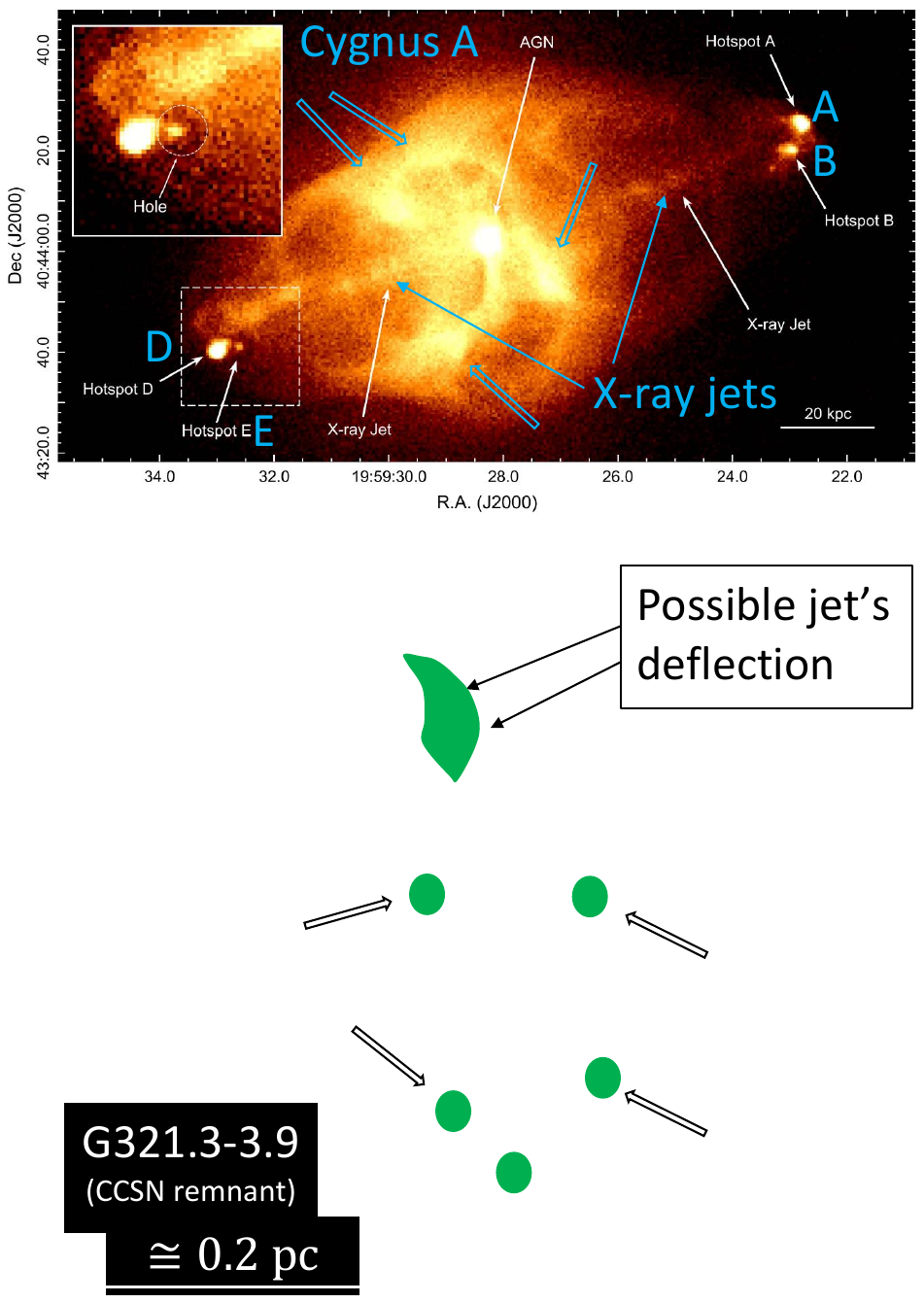}
\end{center}
\caption{Upper panel: An X-ray image of the cluster Cygnus A (0.5-7 keV; from \citealt{Sniosetal2020}). The white marks are in the original figures. The horizontal bar is a scale of 20~kpc.  
Lower pane: Marks of the peaks in 0.7–1.2 keV X-ray emission from an image of the CCSN remnant G321.3–3.9 from \cite{Mantovaninietal2024}. These peaks can be seen much clearer in the original image of \cite{Mantovaninietal2024}, with additional radio emission at 200 MHz that is shown in blue, and X-ray emission (eROSITA all-sky surveys) that is displayed in red (0.2–0.7 keV) and green (0.7–1.2 keV). The horizontal scale is $\simeq 0.2 \pc$ for a distance of $\simeq 1 \kpc$. I added double-line arrows to point at four bright regions in each image. 
}
\label{Fig:CygnusA} 
\end{figure}

I present the X-ray image of the cooling flow cluster Cygnus A that I adapt from \cite{Sniosetal2020} in the upper panel of Figure \ref{Fig:CygnusA}. \cite{Sniosetal2020} suggest that the western jet was deflected from Hotspot B to Hotspot A and the eastern jet was deflected from Hotspot E to Hotspot D. In the lower panel of Figure \ref{Fig:CygnusA} I present the peaks in 0.7–1.2 keV X-ray emission based on an image of the CCSN remnant G321.3–3.9 from \cite{Mantovaninietal2024}. 

I inferred from the morphology of Cygnus A that the northern jet that shaped the X-ray emitting gas in the CCSN remnant G321.3-3.9 was also deflected \citep{ShishkinSoker2024}. While the jets in Cygnus A are still active, the jets that exploded the progenitor of G321.3-3.9 are long gone. 
With double-lined arrows, I also point at four bright X-ray knots in Cygnus A that seem to be part of rings, the ring's edges as projected on the sky plane. These rings are shaped by the jets in Cygnus A. I also point at four bright knots in the image of CCSN remnant G321.3-3.9, possibly belonging to two rings, as in Cygnus A, one in the north and one in the south. Otherwise, two pairs of opposite jets might have shaped the four knots, in addition to the pair of opposite jets launched along the long axis CCSN remnant G321.3-3.9 (for more on this CCSN remnant, see \citealt{ShishkinSoker2024}).      

\begin{figure}
\begin{center}
\includegraphics[trim=0.0cm 0.9cm 0.0cm 3.1cm ,clip, scale=0.40]{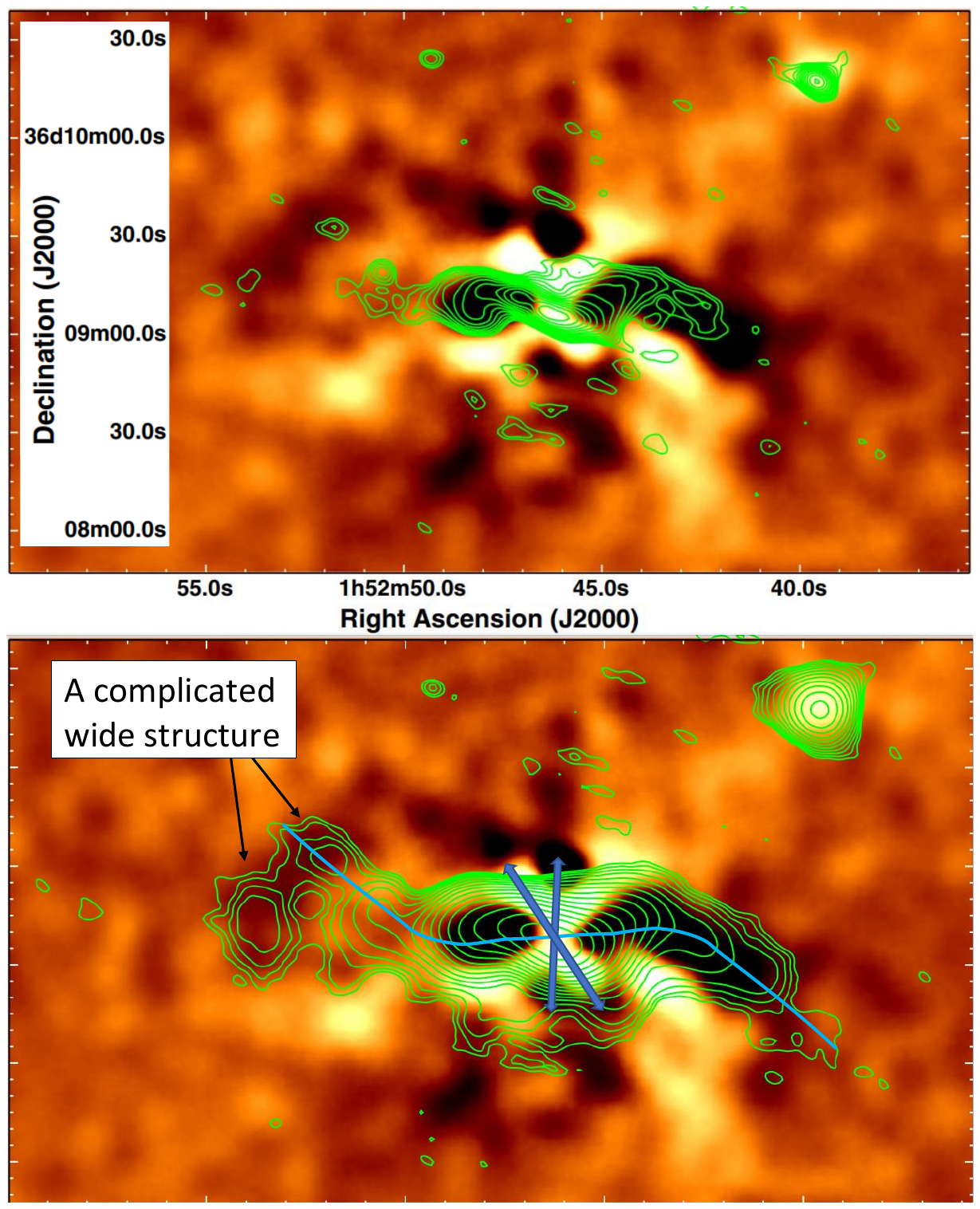}
\end{center}
\caption{Radio contours on top of X-ray residual images of Abell 262 from \cite{Clarkeetal2009}. 
Upper panel: Full resolution GMRT 610 MHz radio contours overlaid on the residual
Chandra X-ray image. 
Lower panel: The radio 610 MHz image is at a lower resolution to be more sensitive to lower surface brightness emission. }
\label{Fig:A262} 
\end{figure}

Another relevant cooling flow is Abell 262, which I present in Figure \ref{Fig:A262}, adapted from \cite{Clarkeetal2009}. The radio contours in the lower panel of Figure \ref{Fig:A262} present an S-shaped morphology and a complicated wide jet ending on the eastern side. This suggests that the CCSN remnant G321.3-3.9 was shaped by both precessing jets, as \cite{ShishkinSoker2024} propose, and by jet deflection on both the south and north sides. There are several cooling flows where observations show precessing jets, e.g., Hydra A (e.g., \citealt{McNamaraetal2000, Nawazetal2016}) and MS 0735.6+7421 (e.g., \citealt{McNamaraetal2005, Vantyghemetal2014}). In the case of AGN jets, though, the precession might be powered by the presence of a secondary supermassive black hole, e.g., as \cite{PizzolatoSoker2005BinaryBH} suggested for MS 0735.6+7421 by noticing its morphological similarity to jet's precession in planetary nebulae known to be shaped by binary interaction.   

Overall, the comparison I present in Figure \ref{Fig:CygnusA} strongly supports that one or three pairs of jets shaped CCSN remnant G321.3-3.9 as suggested by \cite{ShishkinSoker2024}. Since the point-symmetric structure occupies a large volume of the remnant, the jets must have been energetic to the degree that they exploded the massive star progenitor of CCSN remnant G321.3-3.9. 

\subsection{CCSN remnant SNR 0540-69.3} 
\label{subsec:SNR0540-69.3}

\begin{figure}
\begin{center}
\includegraphics[trim=0.0cm 0.8cm 4.0cm 0.0cm ,clip, scale=0.55]{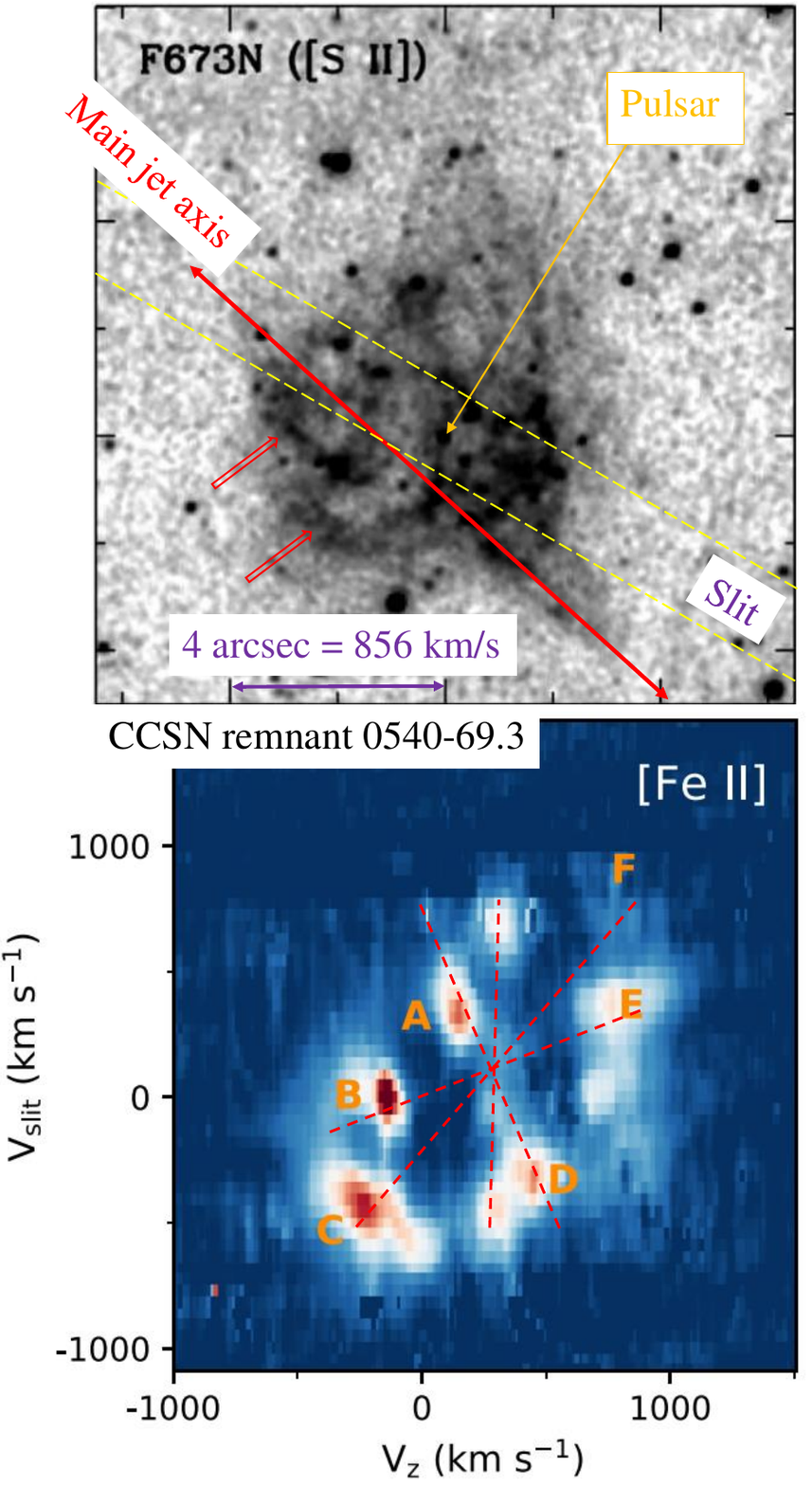}
\end{center}
\caption{The CCSN remnant 0540-69.3 structure in two planes. 
Upper panel: 
An HST observation by \cite{Morseetal2006}. The dashed-yellow lines border the slit position used by \cite{Larssonetal2021} to obtain the Doppler-shift map in the lower panel. The red-double-headed arrow marks the main jet-axis defined by \cite{Soker2022SNR0540}. 
Lower panel: 
A two-dimensional velocity map along the slit between the two dashed-yellow lines on the upper panel (adapted from \citealt{Larssonetal2021}).  Intensity goes from deep blue (low) to deep red (high). The red zones are clumps in the plane of the slit or segments of filaments that cross the plane of the slit.  The velocity along the slit $v_{\rm slit}$ is positive to the northeast, and $v_{\rm z}$ is the velocity along the line of sight. 
The four red-dashed lines are from \cite{Soker2022SNR0540}; they identify the point-symmetric morphology. 
The two panels are of the same scale under the assumption of a remnant age of 1100 yr \citep{Larssonetal2021}. 
}
\label{Fig:SNR0540693} 
\end{figure}

The point-symmetric morphology of CCSN remnant 0540-69.3 appears in its Doppler shift maps of different spectral lines along a slit rather than on the plane of the sky \citep{Larssonetal2021}. 
In  \cite{Soker2022SNR0540}, I identified and analyzed its point-symmetric structure in the frame of the JJEM, including the study of the neutron star kick velocity direction. Here, I limit the presentation to only two maps, an [S \textsc{ii}] image from \cite{Morseetal2006} and the velocity map in the [Fe \textsc{ii}] line as I take from \cite{Larssonetal2021}; these are presented in Figure \ref{Fig:SNR0540693}. The dashed-yellow lines on the upper panel mark the borders of the spectral slit. The two panels have the same scale for an assumed remnant age of 1100 yr \citep{Larssonetal2021}.   

While in \cite{Soker2022SNR0540} I compared the point-symmetric morphology to those of some planetary nebulae thought to be shaped by jets, here I compare the structure of CCSN remnant 0540-69.3 with those of cooling flows and further strengthen the case of shaping by jets. In that paper, I also identified a main jet-axis, marked by the red double-headed arrow on the upper panel of Figure \ref{Fig:SNR0540693}. 

The upper panel of Figure \ref{Fig:SNR0540693} reveals that in addition to the main jet axis, there are also arcs and rings around this axis; I point at two with red double-lined arrows. This structure is similar in several properties, mainly the main jet-axis and rings, to the cooling flow Cygnus A as presented in the upper panel of Figure \ref{Fig:CygnusA}. This similarity strengthens the identification of the main jet axis. I do note that this CCSN remnant has a pulsar wind nebula (e.g., \citealt{Tenhuetal2024}) that also influences its morphology. The relation of the pulsar wind nebula to the shaping by the exploding jets requires further study. 

The main jet axis is close to the spectral strip. As such, the main jet axis might be part of the structure of the velocity map along the slit, as presented in the lower panel of Figure \ref{Fig:SNR0540693}. The velocity map presents pairs of opposite clumps. In \cite{Soker2022SNR0540}, I argued that two to four pairs of jittering jets shaped the inner ejecta in this plane. The cooling flow cluster Abell 2597 (Figure \ref{Fig:A2597}) shows that jets and the bubbles they inflate can indeed shape opposite pairs of dense clumps. This similarity to a cooling flow known to be shaped by jets strengthens the claim that jets shaped the point-symmetric velocity structure of SNR 0540-69.3.

\subsection{CCSN remnant CTB 1} 
\label{subsec:CTB 1}

\begin{figure}
\begin{center}
\includegraphics[trim=0.0cm 5.6cm 12.0cm 0.0cm ,clip, scale=0.78]{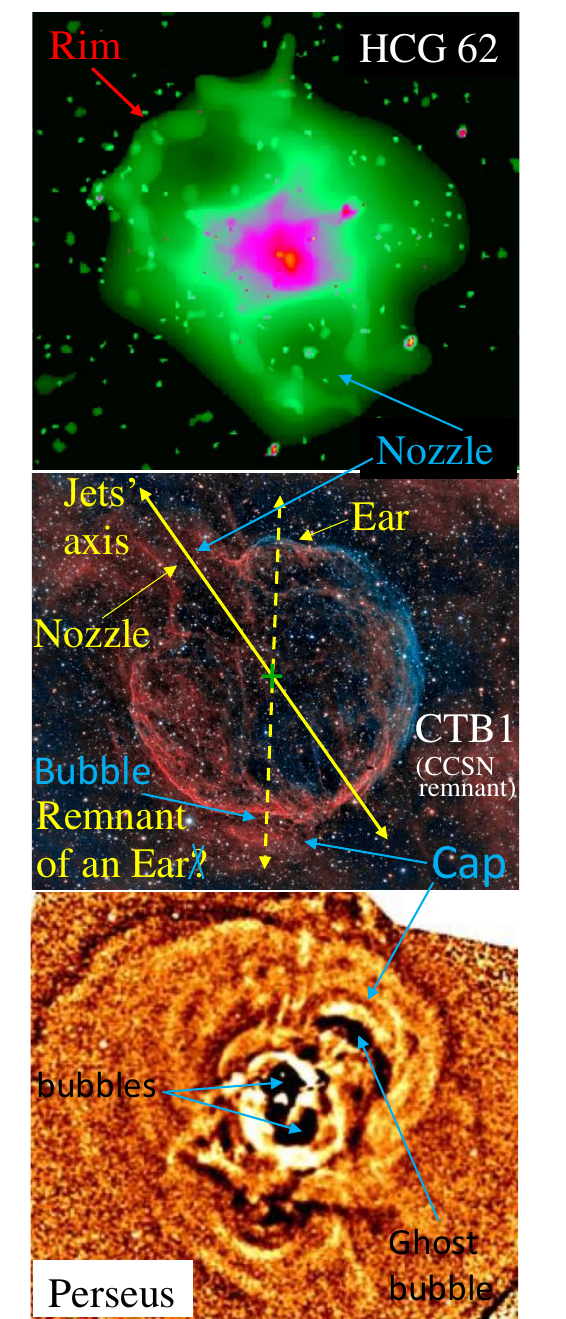}
\end{center}
\caption{Comparing a CCSN remnant (middle panel) with two cooling flows.
Upper panel: A Chandra X-ray image $2^\prime \times 2^\prime = 34 \kpc \times 34 \kpc$ image of HCG 62 galaxy group (Credit: NASA/CfA/J. Vrtilek et al.\footnote{{\tiny https://chandra.harvard.edu/photo/2001/hcg62/}}; see also \citealt{Gittietal2010}). \newline 
Middle panel: CCSN remnant CTB 1 with identification of morphological features by \cite{BearSoker2023CTB1} in yellow, on a picture adapted from Kimberly Sibbald\footnote{{\tiny https://www.flickr.com/photos/spacepaparazzi/51748096139/in/dateposted-public/}} (see also \citealt{ReyesIturbideetal2024} and an image by Tomasz Zwolinski\footnote{{\tiny http://astrozdjecia.pl/}}). Red and blue are for H$\alpha$ and [O \textsc{iii}], respectively.  I define the cap and compare it to that in the Perseus cooling flow cluster in the lower panel.
Lower panel: A Chandra 0.3–7 keV image of the Perseus cooling flow cluster adapted from \cite{Fabianetal2006}. The image is made by subtracting an image smoothed with a Gaussian of dispersion 10 arcsec from one smoothed by 1 arcsec and dividing by the sum of the two images. The horizontal size of the image is $6.5^\prime=143 \kpc$. 
}
\label{Fig:CTB1} 
\end{figure}

\cite{BearSoker2023CTB1} tentatively identified a point-symmetric morphology in the CCSN remnant CTB 1 composed of two jets' axes, as I present in the middle panel of Figure \ref{Fig:CTB1}. While the axis through the nozzle, as marked by the solid-yellow double-headed arrow, is more secure, they consider the axis marked by the yellow-dashed double-headed arrow speculative. By comparing the nozzle to a nozzle in the X-ray image of the galaxy group HCG 62, as I marked in the upper panel, and the `cap' in the southern end of the tentative axis to that in the Perseus cooling flow cluster, as I marked in the lower panel, I strengthen the claim that this CCSN remnant has a point symmetric structure.  
\cite{BearSoker2023CTB1} had a question mark on identifying the southern remnant of an ear. I now remove this question mark.  

The ghost bubble of Perseus is so-called because there is no bright radio inside it, although it was inflated by a jet. The CCSN remnant CTB 1 cap and the cooling flow cluster Perseus cap are elongated tangentially. I, therefore, suggest that the faint region between the main remnant of CTB 1 and the cap was inflated by a jet, and this region might be identified as a ghost bubble. Namely, what \cite{BearSoker2023CTB1} consider to be a tentative identification of an ear, I consider as indeed being a remnant of an ear that was formed by a bubble, as I mark on the figure.  

\subsection{CCSN remnant Vela} 
\label{subsec:Vela}

\begin{figure}
\begin{center}
\includegraphics[trim=0.0cm 11.6cm 10.0cm 0.0cm ,clip, scale=0.78]{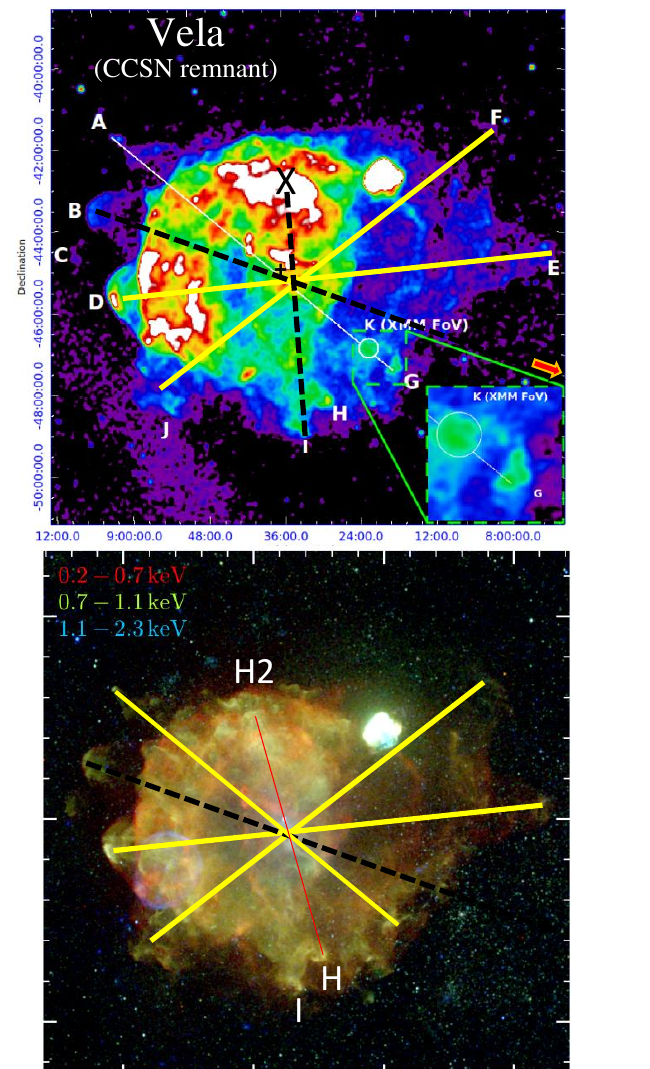}
\end{center}
\caption{Two X-ray images of the Vela CCSN remnant. 
Upper panel: ROSAT X-ray image \citep{Aschenbachetal1995}, based on figure 1 from \cite{Sapienzaetal2021}. The markings of clumps A-F are from \citealt{Aschenbachetal1995}), the white AG-line and the other labeling of the clumps are from \cite{Aschenbachetal1995}. The thick-yellow DE-line,thick-yellow FJ-line, and the two dashed-black lines are additions from \cite{Soker2023Class}. 
Lower panel: Exposure-corrected eROSITA X-ray image color-coded for three bands as indicated in the inset (adapted from  \citealt{Mayeretal2023}, where more details are given). This new image allows the replacement of the speculative X-I-line (vertical black-dashed line) on the Rosat image by a clear point-symmetric feature of clumps H and H2 at equal distances from the center (solid-red line). 
}
\label{Fig:Vela} 
\end{figure}

In \cite{Soker2023Class}, I analyzed the point-symmetric structure of the Vela CCSN remnant and argued that its structure is compatible with the JJEM. In addition to the three pairs of clumps, marked by the white line and the two yellow lines on a ROSAT X-ray image in the upper panel of Figure \ref{Fig:Vela}, I speculated on two more symmetry lines (black-dashed lines). The new eROSITA X-ray image in the lower panel, adapted from \cite{Mayeretal2023},  reveals that there is indeed another pair, but instead of the X-I-line, it is the line connecting H and H2 that I draw in the lower panel of Figure \ref{Fig:Vela}. Note that the center of the red line is at the cross point of the different lines. 

Some clumps, notably B and D, are `caps,' similar to the cap in CCSN remnant CTB 1 and the Perseus cooling flow (Figure \ref{Fig:CTB1}). 
The structure of clumps H and I, as revealed by the new eROSITA image, suggests that the jets were shocked at clump H and then deflected to clump I. This deflection is similar to the deflection in the CCSN remnant G321.3–3.9 and in the cooling flow cluster Cygnus A (Figure \ref{Fig:CygnusA}) I discussed in section \ref{subsec:G321.3–3.9}.

This fourth symmetry line (H-H2 line) that I identify here in Vela and the similarities of some clumps' properties with cooling flows that are known to be shaped by jets (caps and the deflection) make earlier claims that several pairs of jets shaped the point-symmetric morphology of Vela robust. This, in turn, strengthens the JJEM.   

\subsection{CCSN remnant Cassiopeia A} 
\label{subsec:CassiopeiaA}

The new study by \cite{BearSoker2024} shows that the point-symmetry structure of Cassiopeia A is the most robust observational evidence for the JJEM. 

The Cassiopeia A CCSN remnant has been known for a long to have morphological signatures of jets (e.g., \citealt{Reedetal1995, FesenGunderson1996, LamingHwang2003, MilisavljevicFesen2013, FesenMilisavljevic2016, Zhanetal2022, Ikedaetal2022, Kooetal2023, Sakaietal2023,  Milisavljevicetal2024, Vinketal2024}). \cite{PapishSoker2014Planar} speculated that jittering jets in a plane, as they argued to be favored in the JJEM, shaped the ejecta of Cassiopeia A. In an earlier paper 
\citep{Soker2017RAA}, I claimed that the JJEM could better explain the morphology of Cassiopeia A than the delayed neutrino explosion mechanism. However, only in \cite{BearSoker2024} did we reveal the full glory of the point-symmetric morphology of Cassiopeia A and argue that this must have been exploded by jittering jets. According to the delayed neutrino explosion mechanism, the jets are post-explosion processes and do not play a role in the explosion process (e.g., \citealt{Orlandoetal2021}). 

I present two images from \cite{BearSoker2024} who identified several symmetry lines of the point-symmetric morphology and constructed a \textit{point-symmetric wind-rose} of ten lines. The X-ray bright region of Cassiopeia A forms a thick ring (e.g., \citealt{MilisavljevicFesen2013}) that happened to be at a slight angle to the plane of the sky. There are several symmetry axes in that plane, i.e., each axis connects two opposite morphological features of a pair. I emphasize that not all lines correspond to jets. For example, some dense clumps might be formed between jet-inflated bubbles or lobes, as in some cooling flow clusters, e.g., Abell 2052, which I discuss below. In addition to these several symmetry axes, the ring of the bright X-ray zones defines another symmetry axis perpendicular to the ring's plane (more or less along the line of sight.  
\begin{figure}
\begin{center}
\includegraphics[trim=10.6cm 2.1cm 0.0cm 2.7cm ,clip, scale=0.52]{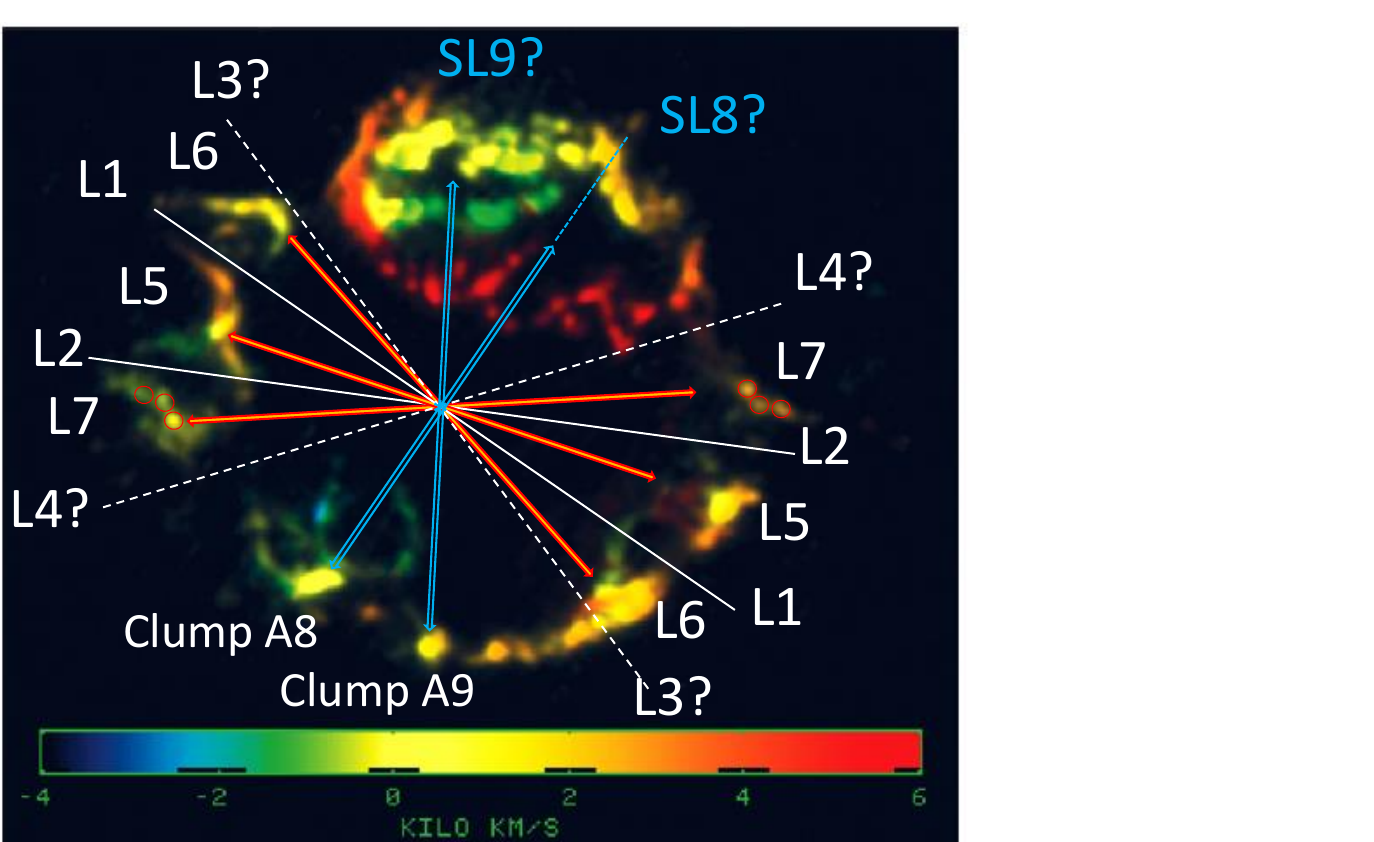}
\includegraphics[trim=5.7cm 3.3cm 1.9cm 0.3cm ,clip, scale=0.42]{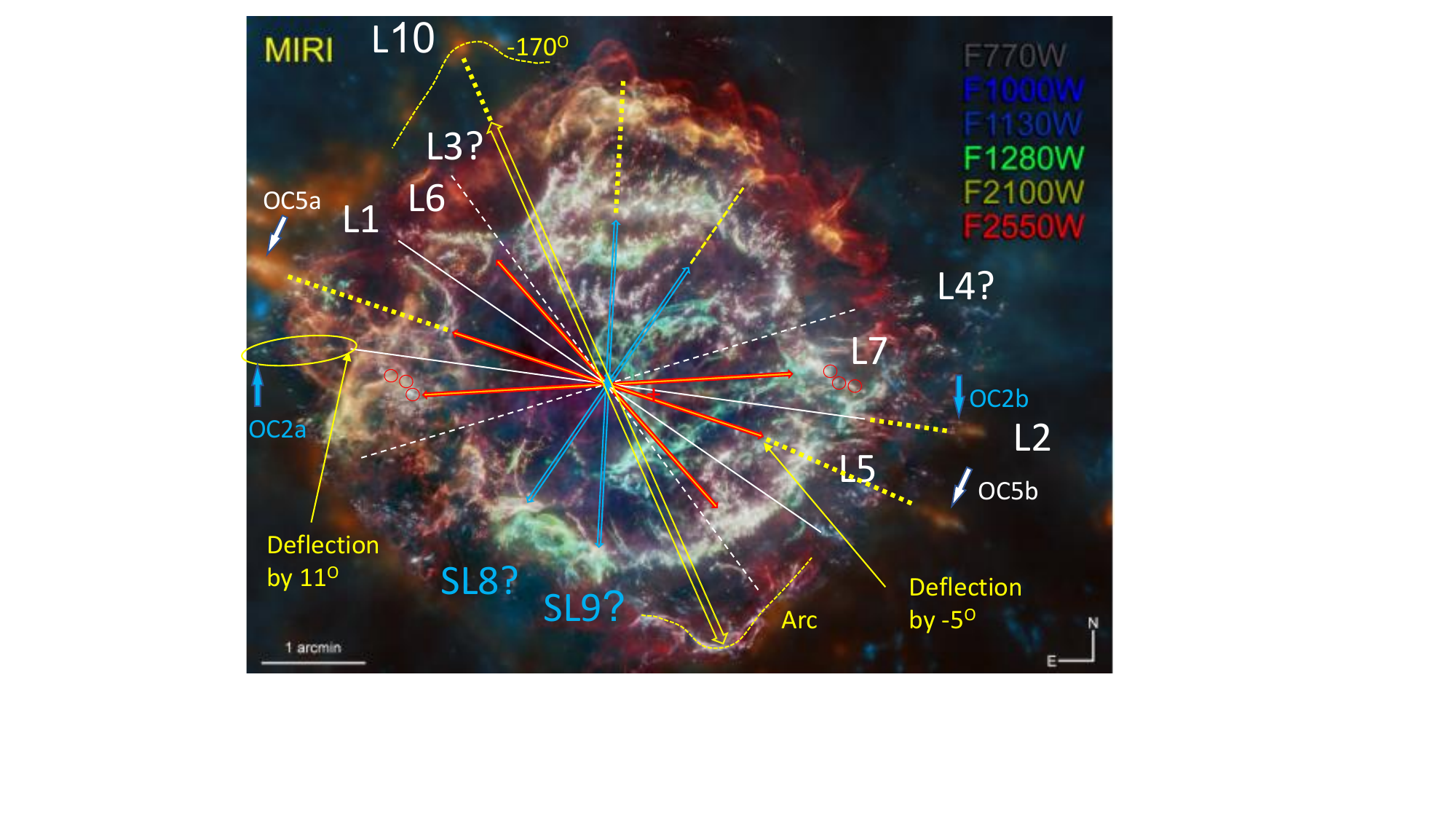}
\end{center}
\caption{Images of CCSN remnant Cassiopeia A (not on the same scale).
Upper panel: A velocity map (by Doppler shift) of Cassiopeia A in the $6.99 \mu$m [Ar \textsc{ii}] line taken from \cite{DeLaneyetal2010}. The identification of symmetry axes and the \textit{point-symmetric wind-rose} they form are from \cite{BearSoker2024}. The velocity scale is from $-4000 \km \s^{-1}$ (deep blue) to $6000 \km \s^{-1}$ (deep red).
Lower Panel:
A JWST image of Cassiopeia A adapted from \cite{Milisavljevicetal2024}. The identifications of point-symmetrical morphological futures and the point-symmetric wind-rose are from \cite{BearSoker2024}. Note the additional symmetry axis L10. The distances of arrow OC2a and arrow OC2b from the cross point of the lines are equal, and so are the distances of arrow OC5a and arrow OC5b.  
The continuation towards OC2a requires a deflection by $11^\circ$ (counterclockwise) and that towards OC5b by $-5^\circ$ (clockwise).   
}
\label{Fig:ArgonMaps1} 
\end{figure}

In the JJEM, the newly born neutron star launches all pairs of jets, several to a few tens, within a very short time, equal to a few times the dynamical time of the pre-collapse stellar core. The total jets' activity phase lasts for a typical time of $\simeq 1-10 \s$, which is $\simeq 3-7$ orders of magnitude shorter than the dynamical time of the exploding star and 9-12 orders of magnitude shorter than the age of observed  CCSN remnants. In cooling flows, the activity phase of jets $\approx 10^7 \yr$ is of the order of the age of observed bubbles or shorter by only about one order of magnitude.  
For that, in CCSN remnants, all pairs of jets are coeval. This is not the case for most jet-inflated bubbles in cooling flows. Almost coeval pairs of bubbles are the two pairs in the cooling flow cluster RBS 797 (e.g., \citealt{Ubertosietal2021, Ubertosietal2023}). The two pairs of bubbles in RBS 797 are almost perpendicular to each other. This rapid change in the jets' axis suggests that the accretion disk and not the spin of the supermassive black hole determines the jets' direction (e.g., \citealt{Soker2022jit}). 
 The reason is that there is not enough angular momentum in the accreted gas between the two jet-launching episodes to change the direction of the spin of the supermassive black hole that launches the jets.  
The hydrodynamical simulations by \cite{Gottliebetal2022} show that the jets are launched along the symmetry axis of the accretion disk rather than the black hole spin.  

\begin{figure}
\begin{center}
\includegraphics[trim=0.0cm 2.2cm 11.0cm 0.0cm ,clip, scale=0.72]{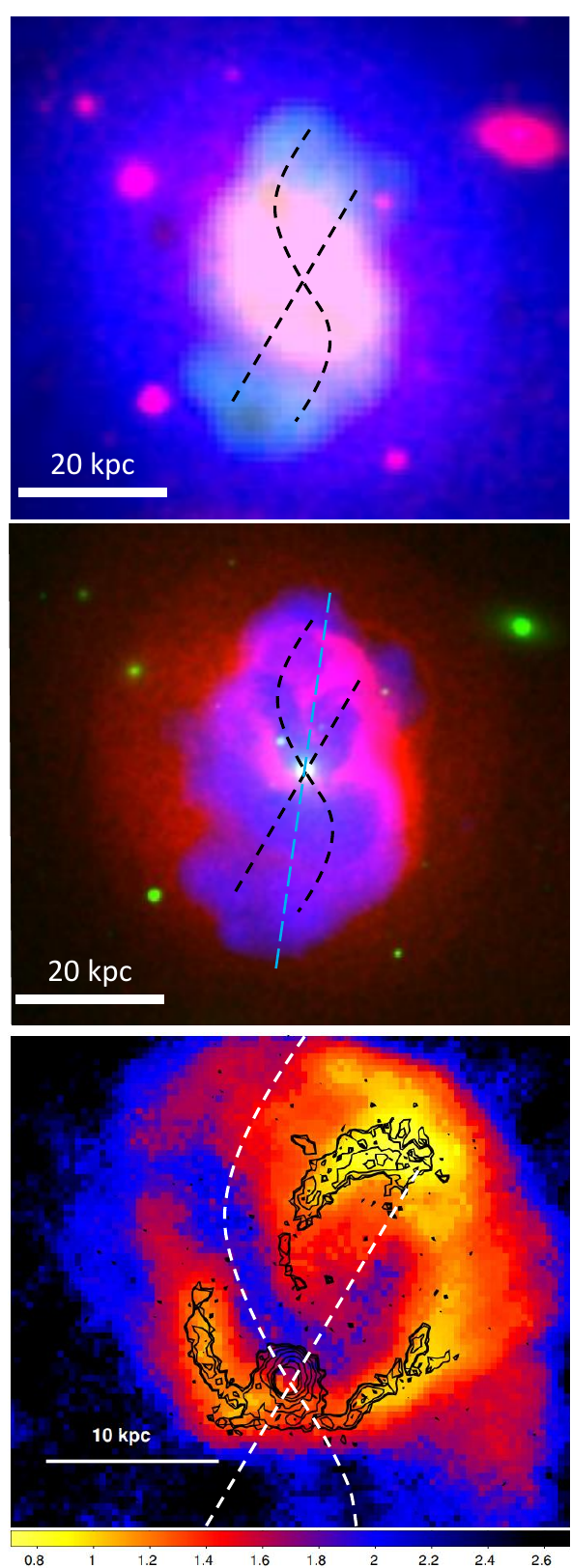}
\end{center}
\caption{Images of Abell 2052. I added the two dashed-black lines to identify a point symmetry.  
Top panel: An image from \cite{Blantonetal2010}: Composite X-ray (blue, Chandra) optical (red, Digitized Sky Survey), radio [green/yellow, 1.4 GHz]. Middle panel: An image from \cite{Blantonetal2011}: Composite Chandra X-ray (red), VLA 4.8 GHz (blue), and a Sloan Digital Sky Survey in r-band (green). The dashed-light-blue line indicates an axis around which I suggest the jets might have precessed. 
Bottom panel: An image from \cite{Blantonetal2011}:
H$\alpha$ contours from \cite{McDonaldetal2010} superposed onto a temperature map; the scale bar is kT in units of keV.  
I suggest that the inner bubbles were shaped by a pair of opposite jets with rapidly varying jet axes, possibly precessing.  
}
\label{Fig:A2052} 
\end{figure}

\cite{Blantonetal2010} and \cite{Blantonetal2011} present deep X-ray images of the cooling flow cluster Abell 2052. I present three images from their studies in Figure \ref{Fig:A2052}. I added the two dashed lines on the three panels, one straight and one S-curved; these two lines' structures are identical in all three panels. Abell 2052 has a tangentially elongated bubble to the south of the center. I suggest that it is composed of two bubbles on the plane of the sky, opposite to the two bubbles on the plane of the sky on the north, which are separated by the cold X-ray gas and the H$\alpha$ filament (lower panel of Figure \ref{Fig:A2052}). If this is indeed the structure, most likely, the inner bubbles of Abell 2052 were shaped by an almost coeval pair of opposite jets with slightly varying axis directions. Most probably, the axis of the jets was precessing around an axis more or less along the H$\alpha$ filament, as I draw by the dashed-light-blue line in the middle panel of Figure \ref{Fig:A2052}. 

 On larger scale than the panels of Figure \ref{Fig:A2052}, \cite{Blantonetal2011} identified a large spiral structure of the cool gas in Abell 2052. The inner part of the large-scale spiral structure is the cool gas in the inner region (yellow and red colors on the bottom panel of Figure \ref{Fig:A2052}). The outer spiral structure does influence the structure of the inner gas. However, the radio emission (middle panel) indicates that the bubbles and filaments also relate to jet activity. I refer to the morphology that the jets shaped. A future quantitative comparison of morphologies will have to take into account the influence of the outer spiral structure as well.

The relevant properties of Abell 2052 are as follows. (1) It has a point-symmetrical morphological component with a highly distorted symmetry, as the southern X-ray deficient bubbles (cavities) are not identical in size and shape to the northern bubbles. (2) There is a dense gas between close bubbles.  
These properties of a morphology that is known to have been shaped by pairs of jets suggest that point-symmetric CCSN remnant morphologies that are not exactly point-symmetric, i.e., two opposite sides are not equal in distance, shape, or size, are also shaped by jets. Also, Abell 2052 and some other cooling flows show that dense clumps and filaments might be compressed gas between jet-inflated bubbles and lobes rather than the tips of lobes. Tips of jet-inflated lobes can also form opposite clumps, as some planetary nebulae and some nebulae around post-asymptotic giant branch stars show (e.g., \citealt{Soker2024SN1987A}).   

\cite{Picquenotetal2021} argued against a jet-driven explosion of Cassiopeia A because opposite red-shifted and blue-shifted features in Cassiopeia A are not precisely at $180^\circ$ to each other. That many pairs of bubbles in cooling flows are not at $180^\circ$ to each other disqualifies the argument by \cite{Picquenotetal2021} against the JJEM.

\subsection{CCSN remnant candidate G107.7-5.1} 
\label{subsec:G107.7-5.1}

\begin{figure}
\begin{center}
\includegraphics[trim=0.3cm 0.0cm 0.0cm 0.0cm ,clip, scale=0.49]{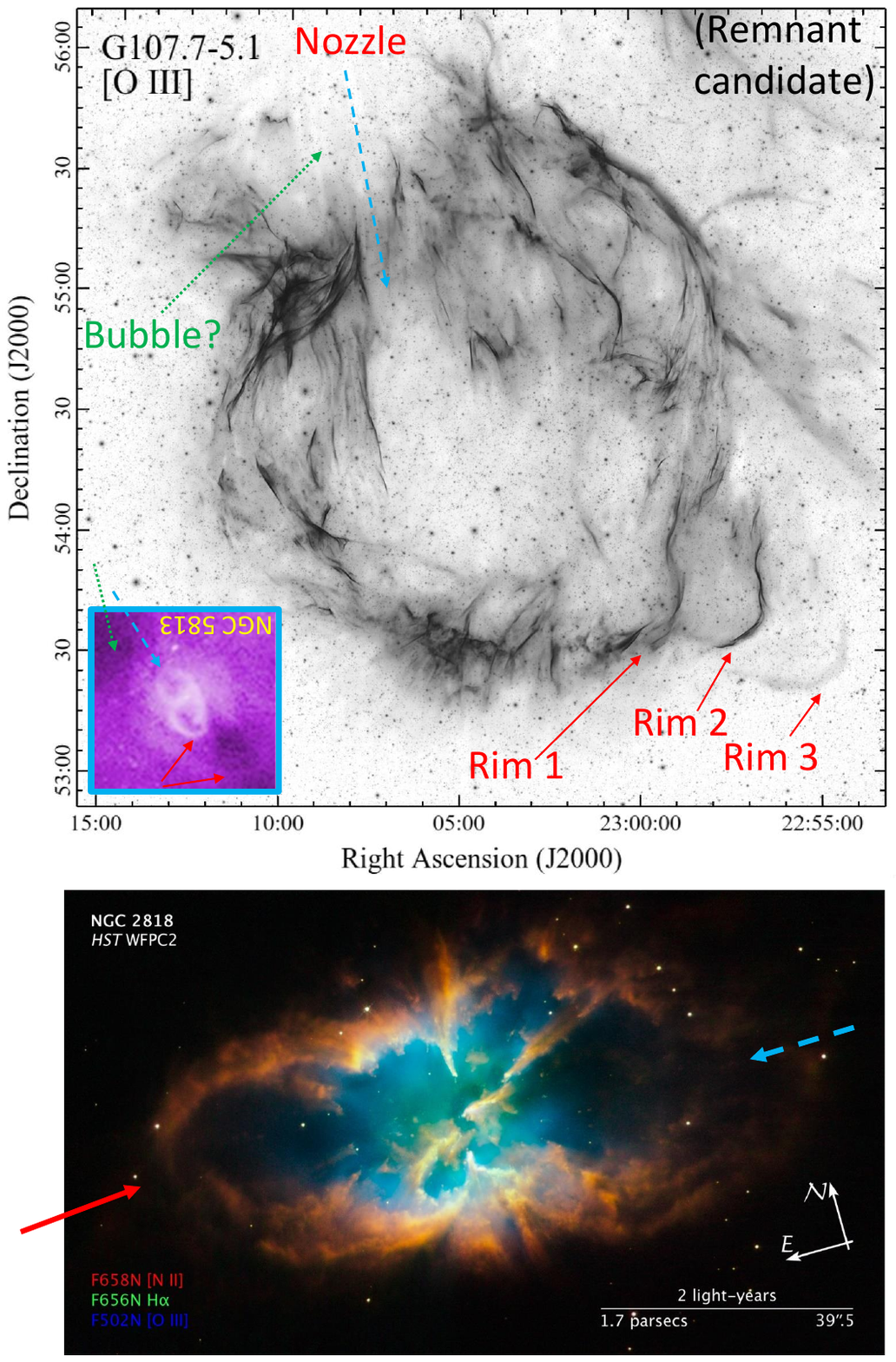}
\end{center}
\caption{Upper panel: An image of a new supernova remnant candidate G107.7-5.1 adapted from \cite{Feseonetal2024}. I added the mark of the rims, the nozzle, and a possible jet-inflated bubble. The inset shows the inner part of the cooling flow NGC 5813 (Figure \ref{Fig:NGC5813}) turned by $180^\circ$, i.e.,  north is down, and east is to the right. 
I use the same arrow keys to point at two rims, a nozzle, and a bubble in NGC 5813. 
The G107.7-5.1 asymmetry of one side of the rims and the other of a nozzle + bubble is similar to the rim-nozzle asymmetry in the cooling flow group NGC 5813, which is known to be shaped by jets. 
Lower panel: An HST image of the bipolar planetary nebula NGC 2818 to further emphasize unequal opposite sides: one has a nozzle (dashed-light-blue arrow) and the other a rim (solid-red arrow). (Credit: NASA, ESA, and the Hubble Heritage Team.)
}
\label{Fig:G107} 
\end{figure}

I end with a more speculative identification of multi-episode jet activity, which demands further study. In the upper panel of Figure \ref{Fig:G107}, I present a new image of a newly identified supernova remnant candidate G107.7-5.1 \citep{Feseonetal2024}.
Due to its highly aspherical morphology, I assume it is a CCSN remnant. I added marks of three rims, a nozzle, and a possible bubble that I identified in this CCSN remnant candidate. The inset is the inner part of the cooling flow NGC 5813 that I present in Figure \ref{Fig:NGC5813}, and turned by $180^\circ$, namely, north is down, and east is to the right.
Due to the important morphology of unequal sides, one with rims and one with a nuzzle, I also present a bipolar planetary nebula in the lower panel. This planetary nebula, NGC 2818, shows a clear nozzle-rim asymmetry (see \citealt{Derlopaetal2024} for a recent detailed study of NGC 2818). I attribute the rim-nozzle morphology to jets with unequal powers. The weaker jet inflates a bubble with its front rim, while the stronger jet breaks out from the main shell/ejecta and opens a nozzle.   

The middle panel of Figure \ref{Fig:NGC5813} shows two inner northeast rims in the cooling flow NGC 5813, namely, 1E and 2E; the two red arrows in the inset of Figure \ref{Fig:G107} point at these two rims. The two other arrows in the inset point at the nozzle and one of the bubbles of NGC 5813. In both supernova remnant candidate G107.7-5.1 and the cooling flow NGC 5813 there is a slight misalignment between consecutive rims. 

If supernova remnant candidate G107.7-5.1 is established as a CCSN remnant, the above similarity suggests that at least three jet-launching episodes left their marks on the remnant. In this case, there were small changes in the directions of the jets' axes in these three jet-launching episodes. 

\begin{figure*}
\begin{center}
\includegraphics[trim=0.0cm 11.1cm 0.0cm 0.0cm ,clip, scale=0.85]{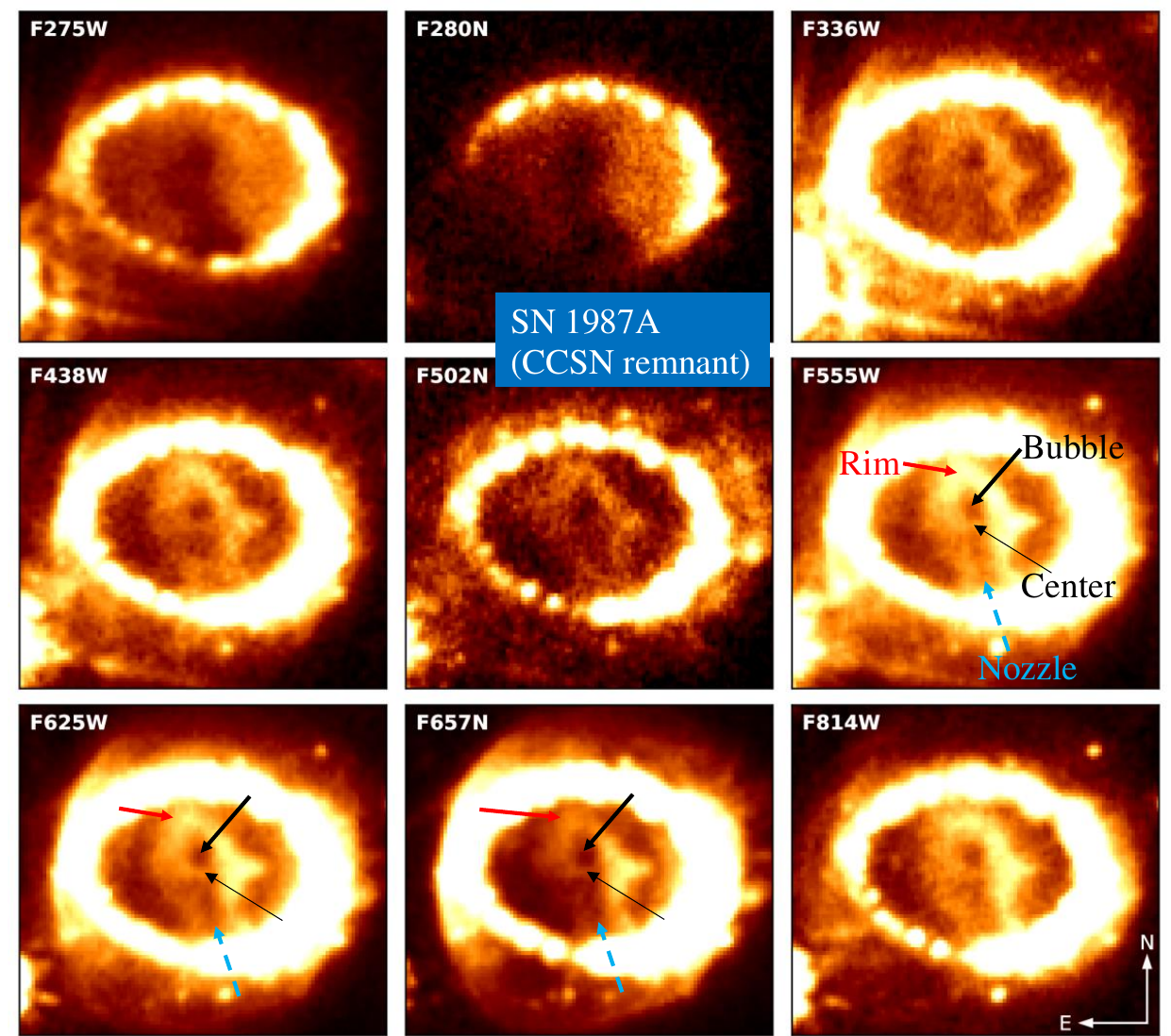} 
\end{center}
\caption{HST/WFC3 images of SN 1987A in nine filters and at 35.5 years post-explosion from \cite{Rosuetal2024}. I identified a northern bubble (thick-black arrow) with its front rim (red arrow) and a southern nozzle (dashed-light-blue arrow). I marked it on three panels, but others also show this asymmetry. The rim-nozzle asymmetry is seen in planetary nebulae and cooling flows (e.g., Figures \ref{Fig:NGC5813}, \ref{Fig:CTB1}, \ref{Fig:G107}), which are known to be shaped by jets. This further strengthens the JJEM for SN 1987A  
}
\label{Fig:SN197AHST}
\end{figure*}

For the importance of the rim-nozzle asymmetry and its ubiquity, I return to the remnant of SN 1987A and present in Figure \ref{Fig:SN197AHST} new HST images of the ejecta of SN 1987A from \cite{Rosuetal2024}, taken 12,980 days after the explosion. I refer to the main elongated structure of the ejecta that still resides inner to the equatorial ring, seen as the over-exposed ellipse in the images. This ejecta region is marked by the two red-dashed lines in Figure \ref{Fig:SN1987A}. I identify in these images a prominent rim-nozzle asymmetry. I marked this asymmetry by the bubble (thick-black arrow) and its front rim (red arrow) on the north and a nuzzle (dashed-light-blue arrow) on the south. The similarity of this rim-nozzle asymmetry with cooling flows and planetary nebulae further supports the shaping by jets scenario for SN 1987A and, by that, a jet-driven explosion, i.e., the JJEM.      

\section{The case of W49B} 
\label{sec:W49B}

\begin{figure}
\begin{center}
\includegraphics[trim=0.0cm 8.0cm 4.9cm 0.0cm ,clip, scale=0.52]{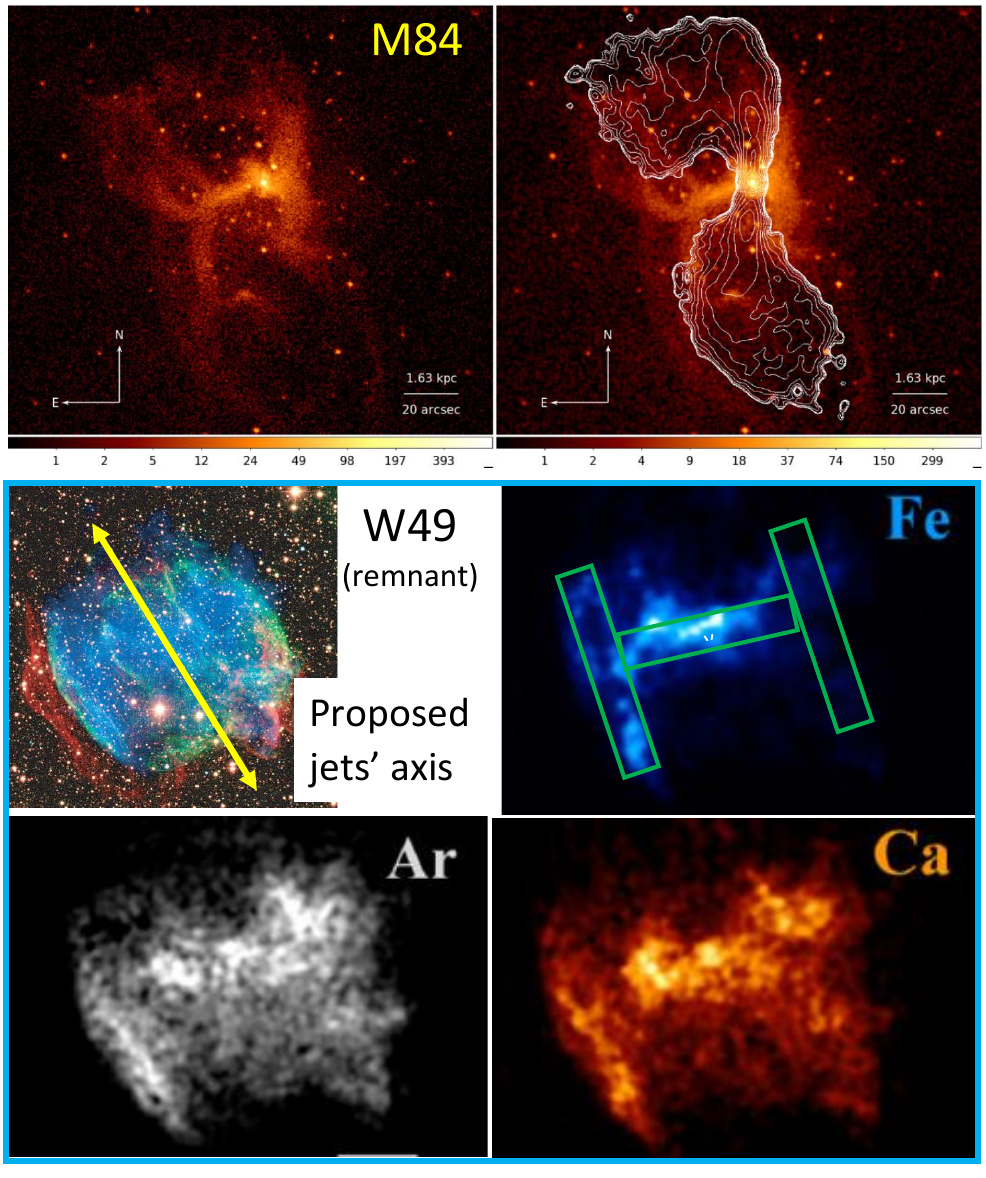}
\end{center}
\caption{Upper two panels: X-ray (0.5–2 keV) image from \cite{Bambicetal2023} of the cooling flow galaxy M84, with 5 GHz radio
contours (white) on the right panel. The colors denote X-ray counts. 
The lower four panels depict the morphology of the supernova remnant W49B.
Middle-left is an image from Chandra's website (http://chandra.harvard.edu/photo/2013/w49b/; Credit: X-ray: NASA/CXC/MIT/L. Lopez et al.; Infrared: Palomar; Radio: NSF/NRAO/VLA), combining X-rays in blue and green (Chandra X-ray Observatory), radio in pink (Very Large Array) and infrared in yellow(Caltech's Palomar Observatory).  The middle-right and lower panels show the X-ray `H-shaped` morphology of W49B in X-ray lines as indicated: Fe~xxv, Ar~xvii and Ar~xviii, and Ca~xix and Ca~xx, respectively.  The later three images are from \cite{Lopezetal2013}; \cite{Lopezetal2013}. 
 The `H-shaped` morphology is marked in green as in \cite{BearSoker2017W49B}.   }
\label{Fig:M84} 
\end{figure}

The supernova remnant W49B is unique (for recent studies of this remnant, see, e.g., \citealt{Siegeletal2020} and \citealt{Satoetal2024}). Its inner part has an H-shaped morphology, similar to some planetary nebulae \citep{BearSoker2017W49B} and to the X-ray morphology of the cooling flow in the galaxy M84, as I present in Figure \ref{Fig:M84}.  
These similarities, and with the two opposite jets in M84, as seen in the radio in Figure \ref{Fig:M84}, imply that the morphology was shaped by jets parallel to the legs of the H-shaped morphology (e.g., \citealt{BearSoker2017W49B,  
GrichenerSoker2023}). The radio jets carved the H-shaped X-ray morphology of M84 (e.g., \citealt{Bambicetal2023}). 

The W49B remnant, however, has a high iron abundance (e.g., \citealt{Siegeletal2020}), unlike CCSNe. This brought \cite{GrichenerSoker2023} to suggest that it was not a CCSN. 
\cite{GrichenerSoker2023}  suggest the common envelope jets supernova scenario for W49B. In this scenario, a neutron star spiraled-in inside the envelope of a red supergiant star and tidally destroyed the massive star's core as it accreted mass from the core. A thermonuclear outburst of the disrupted core partially powered and shaped the ejecta. In addition, the neutron star launched jets that curved the H-shaped morphology. 

The topic of this study is shaping by jets. The new comparison of W49B to the cooling flow galaxy M84 strengthens the suggestion that the H-shaped is due to jets and that their axis was parallel to the legs of the `H' as, e.g.,   \cite{BearSoker2017W49B} and \cite{GrichenerSoker2023} argued. \cite{Lopezetal2013} and \cite{Keohaneetal2007} suggest that the shaping jets were perpendicular to this suggestion; namely, they proposed jets along the horizontal bar of the `H-shaped' morphology. The orientation of the jets that shaped M84 shows that the jets' axis was as argued by \cite{BearSoker2017W49B} and \cite{GrichenerSoker2023}. As with other comparisons in this study, while jets in cooling flows might still be active, in remnants, they are long gone. 

The remnant W49B deserves further study to reveal its true nature. Here, we just strengthened the case for shaping by jets with their axis parallel to the legs of the `H-shaped' remnant.  
 
\section{Summary} 
\label{sec:Summary}

This study takes a new approach to strengthen the arguments of the JJEM that jets shape many CCSN remnants, particularly point-symmetric CCSN remnants. I compared the morphologies of seven point-symmetric CCSN remnants (sections \ref{subsec:N63A} - \ref{subsec:CassiopeiaA}) and one CCSN remnant candidate (section \ref{subsec:G107.7-5.1}) to the X-ray morphologies of several cooling flows. Radio observations show that jets shape point-symmetric structures in cooling flows, e.g., pairs of opposite bubbles (cavities), nozzles, some clumps, and rims. The qualitative similar morphologies I pointed out in this study strongly suggest that jets also shaped the point-symmetric CCSN remnants studied here. The typically large volume of a CCSN remnant shaped by jets further suggests that the energy of the shaping jets is comparable to the ejecta energy. Namely, jets exploded the massive star progenitor.

In section \ref{sec:Feedback}, I discussed some similarities and differences between the jet feedback mechanism in cooling flows and CCSNe. The main difference is that in cooling flows, typically jets are active for a long time, comparable to the age of the bubbles they inflate or not much shorter, while in CCSNe, the jet's activity phase lasts for about 1-10 \s, which is 9-12 orders of magnitude shorter than the age of remnants. Namely, after the jets shape the exploding star during the explosion process, the CCSN remnant experiences a more or less homologous expansion for most of its evolution (depending on the circumstellar and interstellar media). Another difference is that in CCSNe, the two opposite jets are unequal in many jet-launching episodes. Adding the natal kick of the neutron star and large amplitude instabilities during the explosion, the point-symmetric morphologies of CCSN remnants are not pure. Namely, opposite structural features in a pair are likely to differ in one or more of their properties, like shape, size, distance from the center, and even composition. 

The next significant step would be a quantitative comparison of morphologies. Because of the complicated morphologies, I expect that three-dimensional hydrodynamical simulations of shaping by jets will be required to demonstrate that similar initial conditions, but on hugely different scales, can lead to similar morphologies. The dimensionless quantities (namely, ratios of quantities) of the jets and the ambient gas in cooling flows and CCSNe can then be compared. 

The presently studied similar morphologies of eight point-symmetric CCSN remnants to cooling flows add to the similarity of CCSN remnants, and not only point-symmetric ones, to planetary nebulae (section \ref{sec:intro} for references). My view is that the present study solidifies the JJEM as the main explosion mechanism of CCSNe.   

The competing neutrino-driven explosion mechanism cannot explain point-symmetric morphologies in CCSN remnants (despite being more popular in the literature, e.g., some papers supporting or assuming the delayed neutrino explosion mechanism just from the beginning of 2024: \citealt{Andresenetal2024, BoccioliRoberti2024, Burrowsetal2024, GhodlaEldridge2024, JankaKresse2024, Maunderetal2024, Schneideretal2024, WangTBurrows2024}). I consider the identification of point-symmetry in CCSNe, as expected by jet-shaping, to be the most severe challenge to the neutrino-driven explosion mechanism, one that the supporters of the neutrino-driven mechanism have ignored till now. This severe challenge adds to the neutrino-driven explosion mechanism's other challenges. For example, research groups disagree on even the qualitative results, and more so on quantitative results! \footnote{In a review talk in the \textit{Transients Down Under} meeting, held in Melbourne, Australia on 29 January 2024, Hans-Thomas Janka writes in the summary of his talk about the delayed neutrino explosion mechanism:  ``3D models of different groups disagree: there are qualitative and quantitative differences!''.}

I conclude by repeating earlier claims, but in a more vocal voice, that the main explosion mechanism of CCSNe is the JJEM. 

\section*{Acknowledgments}
 
 I thank an anonymous referee for many good comments.  The Pazy Foundation and Israel Science Foundation (769/20) supported this research.

\label{lastpage}
\end{document}